\documentclass[prb,twocolumn,showpacs,preprintnumbers,amsmath,amssymb]{revtex4}

\usepackage{graphicx}
\usepackage{longtable}

\begin{document}

\title{Scheme for adding electron-nucleus cusps to Gaussian orbitals}

\author{A.~Ma, M.~D.~Towler, N.~D.~Drummond, and R.~J.~Needs}

\affiliation{Theory of Condensed Matter Group, Cavendish Laboratory,
University of Cambridge, Madingley Road, Cambridge, CB3 0HE, United
Kingdom}

\date{\today}

\begin{abstract} 
A simple scheme is described for introducing the correct cusps at
nuclei into orbitals obtained from Gaussian basis set electronic
structure calculations.  The scheme is tested with all-electron
variational quantum Monte Carlo (VMC) and diffusion quantum Monte
Carlo (DMC) methods for the Ne atom, the H$_2$ molecule, and
fifty-five molecules from a standard benchmark set. It greatly reduces
the variance of the local energy in all cases and slightly improves
the variational energy. One therefore expects the scheme to yield a
general improvement in the efficiency of all-electron VMC and DMC
calculations using Gaussian basis sets.
\end{abstract}


\maketitle

\section{Introduction}
\label{sec:introduction}

Quantum Monte Carlo (QMC) methods provide a very promising approach
for calculating accurate energies of many-electron systems.  For low
atomic number ($Z$) atoms it is quite common to use all-electron QMC
techniques where every electron is explicitly included in the
simulation, but the computational cost rises rapidly with $Z$. The
scaling behavior can be considerably improved by replacing the core
electrons with pseudopotentials, but this procedure inevitably
introduces errors and it is clearly desirable to perform highly
accurate all-electron QMC calculations for a wider range of atomic
numbers than has been attempted before. In this article we demonstrate
that an accurate representation of the electron--nucleus
cusps~\cite{kato_1957} in the wave function is, not unexpectedly, of
critical importance in such calculations.

The VMC technique and the more accurate DMC
technique~\cite{foulkes_2001} require an approximate many-body trial
wave function, which is normally written as the product of a Slater
determinant, or sum of determinants, and a Jastrow correlation
factor. The quality of the Slater part of the wave function is
extremely important. For small molecules the orbitals are usually
obtained from a single-particle method such as Hartree-Fock (HF)
theory or density-functional theory, or sometimes from a
multi-determinant description such as the Multi-Configuration
Self-Consistent Field (MCSCF) method. Such calculations are normally
performed using standard quantum chemistry packages which use an
atom-centered Gaussian basis.

One of the problems with Gaussian basis sets is that they are unable
to describe the cusps in the single-particle orbitals at the nuclei
that would be present in the exact HF orbitals, because the Gaussian
basis functions have zero gradient at the nuclei on which they are
centered. This can lead to considerable difficulties in QMC
simulations.  In both VMC and DMC methods the energy is calculated as
the average over many points in the electron configuration space of
the local energy, $E_L = \Psi^{-1} \hat{H} \Psi$, where $\hat{H}$ is
the Hamiltonian and $\Psi$ is the many-electron trial wave
function. When an electron approaches a nucleus of charge $Z$ the
potential energy contribution to $E_L$ diverges as $-Z/r$, where $r$
is the distance from the nucleus.\cite{footnote_units} The kinetic
energy operator acting on the cusps in the wave function must
therefore supply an equal and opposite divergence in the local kinetic
energy, because the local energy is constant everywhere in the
configuration space if $\Psi$ is an eigenstate of the
Hamiltonian. Unfortunately, when using orbitals expanded in a Gaussian
basis set, the kinetic energy is finite at the nucleus and therefore
$E_L$ diverges.  In practice one finds that the local energy has wild
oscillations close to the nucleus, which give rise to a large variance
in the energy.  This is undesirable in VMC, but within DMC it can lead
to severe bias and even to catastrophic numerical instabilities.

Within a basis of Slater-type orbitals (STOs) it is possible to
enforce the cusp conditions by imposing constraints on the solutions
of the self-consistent equations~\cite{galek_2005}.  In principle this
appears to be an excellent solution to the cusp problem, but STO codes
will have to be developed much further for this to become a practical
approach for the range of problems we study.  We are interested in
molecular systems, for which we require both single- and
multi-determinant wave functions, and extended systems modeled within
periodic boundary conditions.  We are not aware of STO codes which are
suitable for all of these purposes and, besides, the modifications to
an STO code required to impose the cusp conditions are non-trivial.

It is not in general possible to satisfy the cusp conditions using
STOs in which each basis function is chosen to obey the cusp
conditions at the nucleus on which it is centered, because of the
contributions from the tails of STOs centered on other nuclei.  Manten
and L\"uchow~\cite{manten_2001} have developed a scheme for applying
cusp corrections to Gaussian orbitals in QMC calculations but, as it
similarly relies on correcting individual atom-centered basis
functions, it is not a full solution to the cusp problem.

An alternative solution to the cusp problem might be to enforce the
electron--nucleus cusp condition using the Jastrow factor.  This is
feasible and we have implemented it, but we found it to be
unsatisfactory because a very large number of variable parameters are
required to obtain a good trial wave function.\cite{drummond_2004}

The solution we have adopted in our computer code
\textsc{casino}~\cite{casino} involves the direct modification of the
molecular orbitals so that each of them obeys the cusp condition at
each nucleus.  This ensures that the local energy remains finite
whenever an electron is in the vicinity of a nucleus, although it
generally has a discontinuity at the nucleus.  We apply this
modification to the molecular orbitals, and no alterations to the
Gaussian basis set codes are required.  We note that our algorithm
could also be used for orbitals expanded in other atom-centered basis
sets, such as STOs, again without the need to modify the code which
generated them.

\section{Electron--nucleus cusp corrections}
\label{sec:cusps}

The Kato cusp condition~\cite{kato_1957} applied to an electron at
${\bf r}_i$ and a nucleus of charge $Z$ at the origin is
\begin{equation} 
\left( \frac{\partial \langle\Psi\rangle}{\partial r_{i}}
\right)_{r_{i}=0} = -Z \langle\Psi\rangle_{r_{i}=0}\;, \label{eq:kato}
\end{equation}
where $\langle\Psi\rangle$ is the spherical average of the many-body
wave function about ${\bf r}_{i}=0$.  For a determinant of orbitals to
obey the Kato cusp condition at the nuclei it is sufficient for every
orbital to obey Eq.~(\ref{eq:kato}) at every nucleus.  We need only
correct the orbitals which are non-zero at a particular nucleus
because the others already obey Eq.~(\ref{eq:kato}).  This is
sufficient to guarantee that the local energy is finite at the nucleus
provided at least one orbital is non-zero there.  In the unlikely
case that all of the orbitals are zero at the nucleus then the
probability of an electron being at the nucleus is zero and it is not
important whether $\Psi$ obeys the cusp condition.

An orbital, $\psi$, expanded in a Gaussian basis set can be written as
\begin{equation}
\psi = \phi + \eta \;,
\label{eq:gaussian}
\end{equation}
where $\phi$ is the part of the orbital arising from the $s$-type
Gaussian functions centered on the nucleus in question (which, for
convenience is at ${\bf r} = 0$), and $\eta$ is the rest of the
orbital.  The spherical average of $\psi$ about ${\bf r} = 0$ is given
by
\begin{equation}
\langle{\psi}\rangle = \phi + \langle{\eta}\rangle \;.
\label{eq:spherical_average_gaussian}
\end{equation}
In our scheme we seek a corrected orbital, $\tilde{\psi}$, which
differs from $\psi$ only in the part arising from the $s$-type
Gaussian functions centered on the nucleus, i.e.,
\begin{equation}
\tilde{\psi} = \tilde{\phi} + \eta \;.
\label{eq:corrected_spherical_average_gaussian}
\end{equation}
The correction, $\tilde{\psi}-\psi$, is therefore spherically
symmetric about the nucleus.  We now demand that $\tilde{\psi}$ obeys
the cusp condition at ${\bf r} = 0$,
\begin{equation}
\left( \frac{d \langle\tilde{\psi}\rangle}{d r} \right)_{0} = -Z
\langle\tilde{\psi}\rangle_{0}\;.
\label{eq:cusp_condition}
\end{equation}
Note that $\langle\eta\rangle$ is cusp-less because it arises from
the Gaussian basis functions centered on the origin with non-zero
angular momentum, whose spherical averages are zero, and the tails of
the Gaussian basis functions centered on other sites, which must be
cusp-less at the nucleus in question.  We therefore obtain
\begin{equation}
\left( \frac{d \tilde{\phi}}{d r} \right)_{0} = -Z\left(
\tilde{\phi}(0) + \eta(0) \right) \;.
\label{eq:cusp_corrected}
\end{equation}
We use Eq.~(\ref{eq:cusp_corrected}) as the basis of our scheme for
constructing cusp-corrected orbitals. 


\section{Cusp correction algorithm}
\label{sec:cusp_correction}

One could conceive of correcting the orbitals either by adding a
function to the Gaussian orbital inside some reasonably small radius,
multiplying by a function (e.g., using the Jastrow factor as mentioned
in Sec.~\ref{sec:introduction}), or by replacing the orbital near the
nucleus by a function which obeys the cusp condition.  However, as the
local energy obtained from Gaussian orbitals shows wild oscillations
close to the nucleus, the best option seems to be the latter one:
replacement of the orbital inside some small radius by a well-behaved
form.

We apply a cusp correction to each orbital at each nucleus at which it
is non-zero.  Inside some cusp correction radius $r_c$ we replace
$\phi$, the part of the orbital arising from $s$-type Gaussian
functions centered on the nucleus in question, by
\begin{equation}
\label{eq:wave_function}
\tilde{\phi} = C + {\rm sgn[\tilde{\phi}(0)]} \exp[p(r)] = C + R(r).
\end{equation}
In this expression ${\rm sgn[\tilde{\phi}(0)]}$ is $\pm1$, reflecting
the sign of $\tilde{\phi}$ at the nucleus, and $C$ is a shift chosen
so that $\tilde{\phi} - C$ is of one sign within $r_c$. This shift is
necessary since the uncorrected $s$-part of the orbital $\phi$ may
have a node where it changes sign inside the cusp correction radius,
and we wish to replace $\phi$ by an exponential function, which is
necessarily of one sign everywhere. The polynomial $p$ is given by
\begin{equation}
p = \alpha_0 + \alpha_1r + \alpha_2r^2 + \alpha_3r^3 + \alpha_4 r^4
\;,
\end{equation}
and we determine $\alpha_0$, $\alpha_1$, $\alpha_2$, $\alpha_3$, and
$\alpha_4$ by imposing five constraints on $\tilde{\phi}$.  We demand
that the value and the first and second derivatives of $\tilde{\phi}$
match those of the $s$-part of the Gaussian orbital at $r=r_c$. We
also require that the cusp condition is satisfied at $r=0$.  We use
the final degree of freedom to optimize the behavior of the local
energy in a manner to be described below.  However, if we impose such
a constraint directly the equations satisfied by the $\alpha_i$ cannot
be solved analytically.  This is inconvenient and we found that a
superior algorithm was obtained by imposing a fifth constraint which
allows the equations to be solved analytically, and then searching
over the value of the fifth constraint for a ``good solution''.  To
this end we chose to constrain the value of $\tilde{\phi}(0)$.  With
these constraints we have:
\begin{enumerate}
\item
\begin{equation}
\ln |\tilde{\phi}(r_c) - C| = p(r_c) = X_1;
\end{equation}
\item
\begin{equation}
\left.\frac{1}{R(r_c)}\frac{d \tilde{\phi}}{d r}\right|_{r_c} =
p'(r_c) = X_2;
\end{equation}
\item
\begin{equation}
\left.\frac{1}{R(r_c)}\frac{d^2 \tilde{\phi}}{d r^2}\right|_{r_c} =
p''(r_c) + p'^2(r_c) = X_3;
\end{equation}
\item
\begin{equation}
\left.\frac{1}{R(0)}\frac{d \tilde{\phi}}{d r}\right|_{0} = p'(0) =
-Z\left(\frac{C + R(0) + \eta(0)}{R(0)}\right) = X_4;
\label{eq:x4}
\end{equation}
\item
\begin{equation}
\ln |\tilde{\phi}(0) - C| = p(0) = X_5.
\end{equation}
\end{enumerate}

\noindent Although the constraint equations are non-linear, they can
be solved analytically, giving
\begin{eqnarray}
\alpha_0 & = & X_5 \nonumber \\
\alpha_1 & = & X_4 \nonumber \\
\alpha_2 & = &  6\frac{X_1}{r_c^2} - 3\frac{X_2}{r_c}   + \frac{X_3}{2}      - 3\frac{X_4}{r_c}   - 6\frac{X_5}{r_c^2} - \frac{X_2^2}{2} \nonumber \\
\alpha_3 & = & -8\frac{X_1}{r_c^3} + 5\frac{X_2}{r_c^2} - \frac{X_3}{r_c}    + 3\frac{X_4}{r_c^2} + 8\frac{X_5}{r_c^3} + \frac{X_2^2}{r_c} \nonumber \\
\alpha_4 & = &  3\frac{X_1}{r_c^4} - 2\frac{X_2}{r_c^3} + \frac{X_3}{2r_c^2} -  \frac{X_4}{r_c^3} - 3\frac{X_5}{r_c^4} - \frac{X_2^2}{2r_c^2}. \label{eq:solution}
\end{eqnarray}
\noindent Our procedure is to solve Eq.~(\ref{eq:solution}) using an
initial value of $\tilde{\phi}(0) = {\phi}(0)$. We then vary
$\tilde{\phi}(0)$ so that the ``effective one-electron local energy'',
\begin{eqnarray}
\label{eq:local_energy}
E_L^s(r) & = & \tilde{\phi}^{-1} \left[ -\frac{1}{2} \nabla^2 -
\frac{Z_{\rm eff}}{r} \right] \tilde{\phi} \\ & = & \nonumber
-\frac{1}{2} \frac{R(r)}{C+R(r)} \left[ \frac{2p'(r)}{r} + p''(r) +
p'^2(r)\right] - \frac{Z_{\rm eff}}{r},
\end{eqnarray}
is well-behaved. Here the effective nuclear charge $Z_{\rm eff}$ is
given by
\begin{equation}
\label{eq:z_eff}
Z_{\rm eff} = Z \left(1 + \frac{\eta(0)}{C + R(0)}\right),
\end{equation}
which ensures that $E_L^s(0)$ is finite when the cusp condition of
Eq.~(\ref{eq:x4}) is satisfied.

We studied the effective one-electron local energies obtained using
Eq.~(\ref{eq:local_energy}) with $Z_{\rm eff}=Z$ for the $1s$ and $2s$
all-electron Hartree-Fock orbitals of neutral atoms calculated by
numerical integration on fine radial grids for atoms up to $Z=82$.  We
noticed that the quantity $E_L^s(r)/Z^2$ is only weakly dependent on
$Z$ in the range $r < 1.5/Z$.  We therefore chose an ``ideal''
effective one-electron local energy curve given by
\begin{eqnarray}
\label{eq:ideal}
\nonumber
\frac{E_L^{\rm ideal}(r)}{Z^2} &=& \beta_0 + \beta_1r^2 + \beta_2r^3 +
\beta_3r^4 \\ & & + \beta_4r^5 + \beta_5r^6 + \beta_6 6r^7 
+ \beta_7r^8.
\end{eqnarray}
The values chosen for the coefficients were $\beta_1 = 3.25819$,
$\beta_2 = -15.0126$, $\beta_3 = 33.7308$, $\beta_4 = -42.8705$,
$\beta_5 = 31.2276$, $\beta_6 = -12.1316$, $\beta_7 = 1.94692$,
obtained by fitting to the data for the $1s$ orbital of the carbon
atom.  The value of $\beta_0$ depends on the particular atom and its
environment.  The ideal effective one-electron local energy for a
particular orbital is chosen to have the functional form of $E_L^{\rm
ideal}(r)$, but with the constant value $\beta_0$ chosen so that the
effective one-electron local energy is continuous at $r_c$.  Hydrogen
is treated as a special case as the 1$s$ orbital of the isolated atom
is only half-filled, and we use $E_L^{\rm ideal}(r) = \beta_0$.

We wish to choose $\tilde{\phi}(0)$ so that $E_L^s(r)$ is as close as
possible to $E_L^{\rm ideal}(r)$ for $0 < r < r_c$, i.e., the
effective one-electron local energy is required to follow the
``ideal'' curve as closely as possible.  In our current implementation
we find the best $\tilde{\phi}(0)$ by minimizing the maximum square
deviation from the ideal energy, $[E_L^s(r) - E_L^{\rm ideal}(r)]^2$,
within this range. Beginning with $\tilde{\phi}(0) = {\phi}(0)$, we
first bracket the minimum then refine $\tilde{\phi}(0)$ using a simple
golden section search. In principle we are more interested in
$E_L^s(r)$ being close to $E_L^{\rm ideal}(r)$ near $r_c$ than near
zero because the probability of an electron being near $r_c$ is
normally much greater than it being near the nucleus. One might
therefore consider using a weighting factor and minimizing, e.g.,
$[r(E_L^s(r) - E_L^{\rm ideal}(r))]^2$. In practical calculations this
was found not to improve the result in general and weighting factors
were not used in our final implementation.

It is clearly important to find an automatic procedure for choosing
appropriate values of the cusp correction radii. Although the final
quality of the wave function in QMC calculations is expected to have
only a relatively weak dependence on its precise value, the optimal
cusp correction radius $r_c$ for each orbital and nucleus should
depend on the quality of the basis set and on the shape of the orbital
in question.  In particular one would expect the cusp correction radii
to become smaller as the quality of the basis set is
improved. Although clearly many other schemes are possible, we choose
the $r_c$ in our implementation as follows.  The maximum possible cusp
correction radius is taken to be $r_{c,max} = 1/Z$.  The actual value
of $r_c$ is then determined by a universal parameter $c_c$ for which a
default value of 50 was found to be reasonable.  The cusp correction
radius $r_c$ for each orbital and nucleus is set equal to the largest
radius less than $r_{c,max}$ at which the deviation of the effective
one-electron local energy calculated with $\phi$ from the ideal curve
has a magnitude greater than $Z^2/c_c$. Appropriate polynomial
coefficients $\alpha_i$ and the resulting maximum deviation of the
effective one-electron local energy from the ideal curve are then
calculated for this $r_c$. As a final refinement one might then allow
the code to vary $r_c$ over a relatively small range centered on the
initial value, recomputing the optimal polynomial cusp correction at
each radius, in order to optimize further the behavior of the
effective one-electron local energy. This is done by default in our
implementation.

When a Gaussian orbital can be readily identified as, for example, a
$1s$ orbital, it generally does not have a node within $r_{c,max}$. In
many cases, however, some of the molecular orbitals have small
$s$-components which may have nodes close to the nucleus. The possible
presence of nodes inside the cusp correction radius complicates the
procedure because the effective one-electron local energy diverges
there. One could simply force the cusp correction radius to be less
than the radius of the node closest to the nucleus, but in practice
nodes can be very close to the nucleus and such a constraint severely
restricts the flexibility of the algorithm. In practice we define
small regions around each node where the effective one-electron local
energies are not taken into account during the minimization, and from
which the cusp correction radius is excluded.

\section{Results}
\label{sec:results}

The above procedure has been implemented in the
\textsc{casino}~\cite{casino} code for both finite systems and systems
periodic in one, two and three dimensions where the orbitals are
represented in Gaussian basis sets.  The code is capable of using
other kinds of basis set including plane waves and a local spline
re-expansion of plane wave orbitals\cite{alfe_2004}, but in such cases
one uses pseudopotentials. Generally these can be forced to be finite
at the nucleus~\cite{trail_2005} and therefore do not lead to cusps in
the orbitals.  Some pseudopotentials however, such as those of the
Stuttgart group~\cite{igel-mann_1988}, diverge like $-1/r$ at the
nucleus. As calculations with these pseudopotentials are normally
performed with Gaussian basis sets, our cusp correction scheme could
in principle be employed to improve the behavior of the local energy
in QMC applications that use them.

In terms of performance, one finds in practice that the set-up
procedure for calculating the optimum cusp parameters before the main
QMC calculation starts takes a negligible amount of CPU time -- at
most a few seconds for large systems. For atoms and molecules the main
orbital evaluation routine is slowed by a few per cent when
calculating the cusp corrections. This increases to around ten per
cent in the periodic case, which is acceptable given the improved
stability and the reduction in the variance of the local energy
obtained in all-electron calculations.

To illustrate the improved capabilities of the code, we have performed
test calculations on the Ne atom, the H$_2$ molecule, and fifty-five
molecules of a standard test set, which will now be described.

\subsection{The Ne atom}
\label{subsec:Ne}

\begin{figure*}
\begin{center}
\includegraphics[width=14cm,clip]{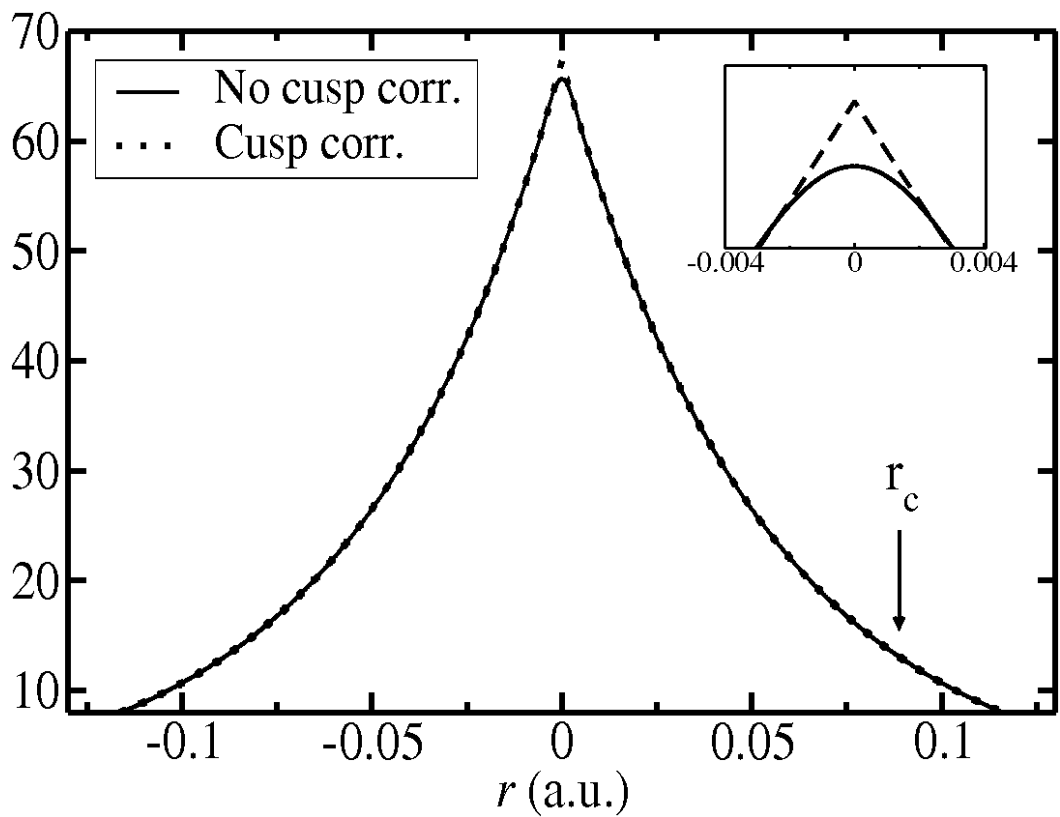}
\caption{The $1s$ orbital of the Ne atom expanded in a Gaussian basis
set with and without the cusp correction.
\label{fig:1s_orb.eps}}
\end{center}
\end{figure*}

In Fig.~\ref{fig:1s_orb.eps} we plot the $1s$ orbital of the Ne atom
with and without the cusp correction. The HF calculations were
performed using the \textsc{crystal} code~\cite{crystal98} with a
reasonably good Gaussian basis set composed of 1 contracted $s$
Gaussian of 6 primitives, 6 uncontracted $s$ functions, and 6
uncontracted $p$ functions, the exponents and contraction coefficients
of which were optimized to minimize the energy. This basis gives a
ground state HF energy of $-128.538450$\,a.u., which is only slightly
higher than the exact HF energy of $-128.547098$\,a.u.

The $1s$ cusp correction radius calculated using the scheme outlined
above is $r_c = 0.0875$\,a.u. This is a little less than the size of
the Bohr radius for the 1$s$ orbital of Ne ($1/Z = 0.1$\,a.u.), but
the constraints at $r=r_c$ and $r=0$ ensure that the corrected orbital
does not deviate much from the original Gaussian orbital except close
to $r=0$.  The inset in Fig.~\ref{fig:1s_orb.eps} shows the behavior
near the nucleus; the cusp in the corrected orbital is readily
apparent.

\begin{figure*}
\begin{center}
\includegraphics[width=14cm,clip]{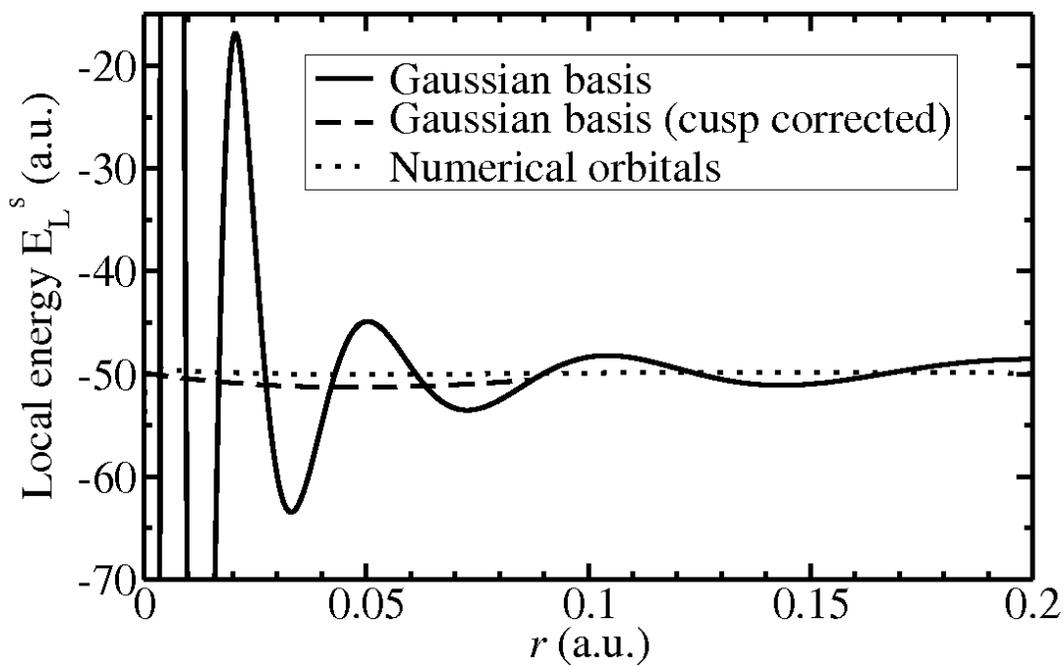}
\caption{The effective one-electron local energy, $E^{s}_L$, versus
distance from the nucleus for the $1s$ orbital of Ne. Data for a
quasi-exact numerical orbital, and the Gaussian orbital with and
without the cusp correction.
\label{fig:ne_1ele.eps}}
\end{center}
\end{figure*}

The effective one-electron local energy of Eq.~(\ref{eq:local_energy})
is plotted as a function of distance from the nucleus in
Fig.~\ref{fig:ne_1ele.eps} for the uncorrected Gaussian orbital, the
cusp-corrected orbital, and for a quasi-exact numerical HF orbital.
The effective one-electron local energy for the exact HF orbital
remains well-behaved over the entire range. The effective one-electron
local energy for the uncorrected Gaussian orbital oscillates far from
the nucleus, and the magnitude of the oscillations grows rapidly at
small $r$, where it reaches a maximum positive value of about
$280$\,a.u., and then tends to $-\infty$ at $r=0$.  The effective
one-electron local energy from the cusp-corrected orbital follows the
uncorrected one from large $r$ down to $r=r_c$, where its gradient
changes abruptly and it begins to approximate the form for the exact
orbital rather closely.

\begin{figure*}
\begin{center}
\includegraphics[width=14cm,clip]{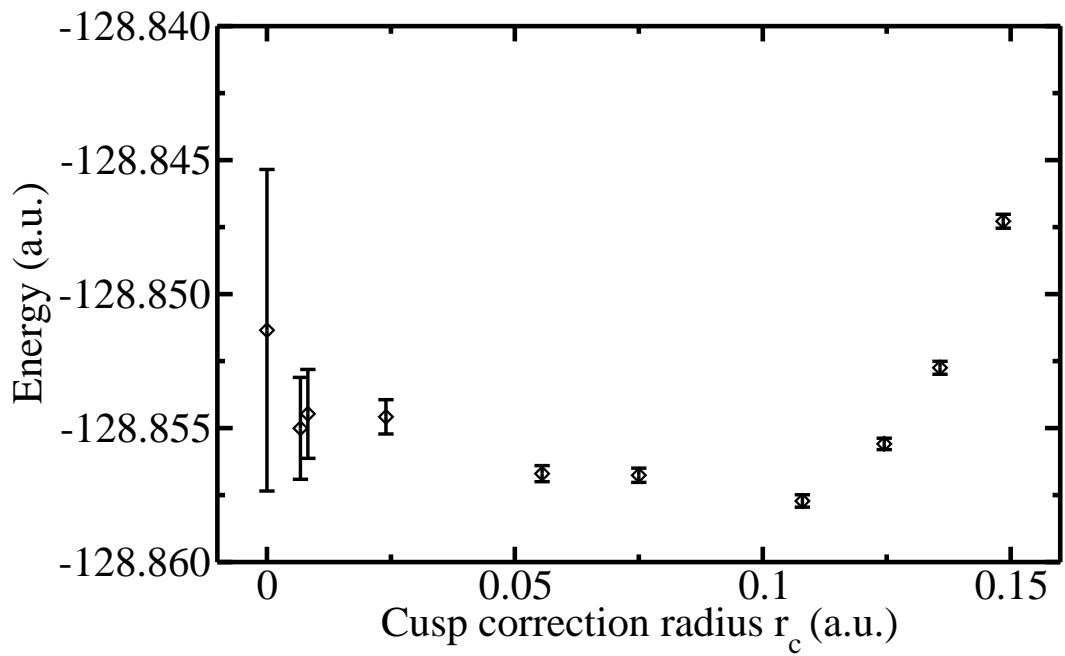}
\caption{The VMC energy of Ne obtained with Slater-Jastrow wave
functions versus the cusp correction radius of the $1s$ orbital. The
length of the error bars is twice the standard error in the mean.
\label{fig:evmc_final.eps}}
\end{center}
\end{figure*}

We tested the cusp-corrected wave functions within VMC and DMC
calculations using the \textsc{casino} code.\cite{casino} First we
performed VMC calculations for Ne with Slater-Jastrow wave functions
including the cusp-corrected orbitals for different values of $r_c$.
The Jastrow factor~\cite{drummond_2004} contained forty-four variable
parameters, whose optimal values were determined separately at each
value of $r_c$ by minimizing the variance of the
energy.\cite{umrigar_1988,kent_1999} In Fig.~\ref{fig:evmc_final.eps}
we plot the VMC energy including statistical error bars versus the
cusp correction radius of the 1$s$ orbital.  For $r_c = 0$ (equivalent
to no cusp correction) the error bar is very large. As the cusp
correction radius is increased it is apparent that the error bar on
the energy is greatly reduced, and that the variational energy itself
is slightly lowered for $r_c < 0.12$\,a.u.  For $r_c > 0.12$\,a.u. the
VMC energy begins to increase, although the variance is still quite
small.  These results indicate that the absence of the cusps in
orbitals expanded in Gaussian basis sets is the largest source of
variance in the energy, and that the cusp correction has removed this
source of variance and has improved the overall quality of the wave
function.  We also note that the results are not very sensitive to
$r_c$, with values between $0.05$\,a.u. and $0.1$\,a.u. giving almost
the same results.  This is important because it suggests that schemes
for choosing $r_c$ automatically, such as the one presented in
Sec.~\ref{sec:cusp_correction}, can be successful.

\begin{figure*}
\begin{center}
\includegraphics[width=14cm,clip]{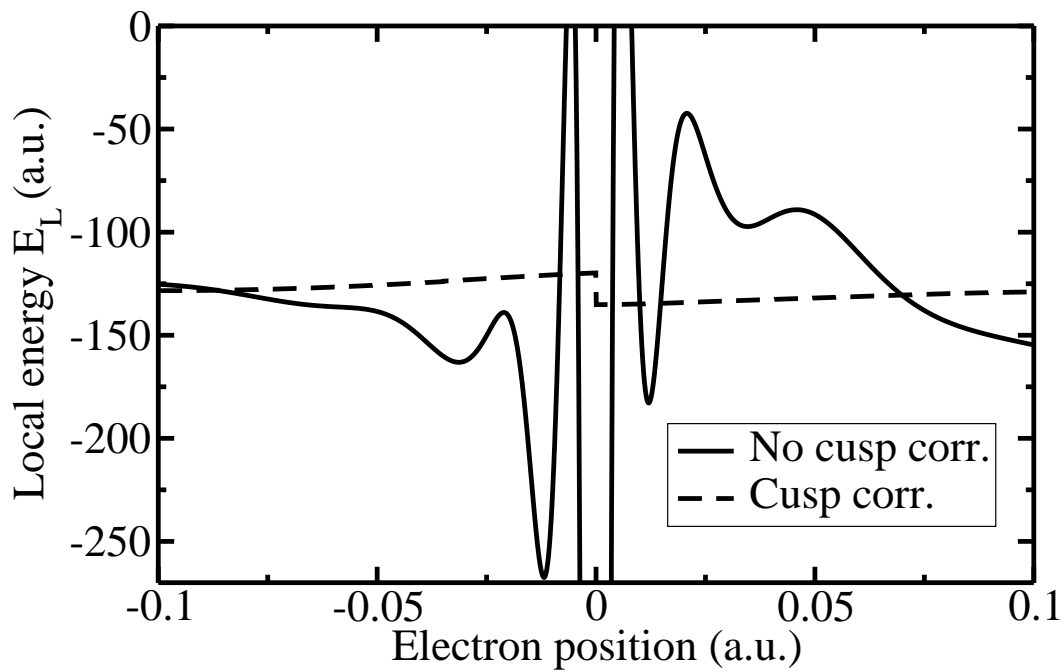}
\caption{The variation of the local energy, $E_L$, as an electron is
moved through the nucleus of a Ne atom which is at the origin.
Slater-Jastrow wave functions are used, both with and without the cusp
correction.
\label{fig:EL_Ne.eps}}
\end{center}
\end{figure*}

In Fig.~\ref{fig:EL_Ne.eps} we show the local energy of Ne, calculated
with the full many-body Hamiltonian, as a function of the separation
of an electron from the nucleus.  This plot was generated by taking an
electron configuration from a VMC run and then calculating the local
energy as the electron closest to the nucleus was moved in a straight
line through the nucleus.  When the cusp correction is included the
local energy is seen to be finite at the origin (but with a finite
discontinuity whose magnitude depends on the positions of all the
electrons).  The local energy for the cusp-corrected wave function
never strays very far from the value which it would have for the exact
wave function.  When the cusp correction is not imposed the local
energy shows wild oscillations of similar magnitude to those of the
effective one-electron local energy in Fig.~\ref{fig:ne_1ele.eps}.

We also tested the cusp-corrected wave function in DMC calculations
and we obtained a DMC energy (extrapolated to zero time step) of
$-128.9218(2)$\,a.u.  This energy is significantly higher than the
exact (non-relativistic and infinite-nuclear-mass) energy of
$-128.9376$\,a.u.~\cite{davidson_1991,chakravorty_1993} due to the use
of the fixed-node approximation, but it is close to the value of
$-128.9238(7)$\,a.u. that we obtained within DMC using quasi-exact
numerical HF orbitals.\cite{drummond_2004}

In order to investigate the range of atomic numbers for which
converged all-electron DMC calculations can feasibly be performed, we
have also calculated the total energies of the noble gas atoms Ar, Kr
and Xe ($Z=18$, $36$, $54$). Details of these calculations, together
with an analysis of the practical scaling behavior of the CPU time
with $Z$, will be given in a separate publication.\cite{ma2}

\subsection{The H$_2$ molecule}
\label{subsec:H2}

\begin{figure*}
\begin{center}
\includegraphics[width=14cm,clip]{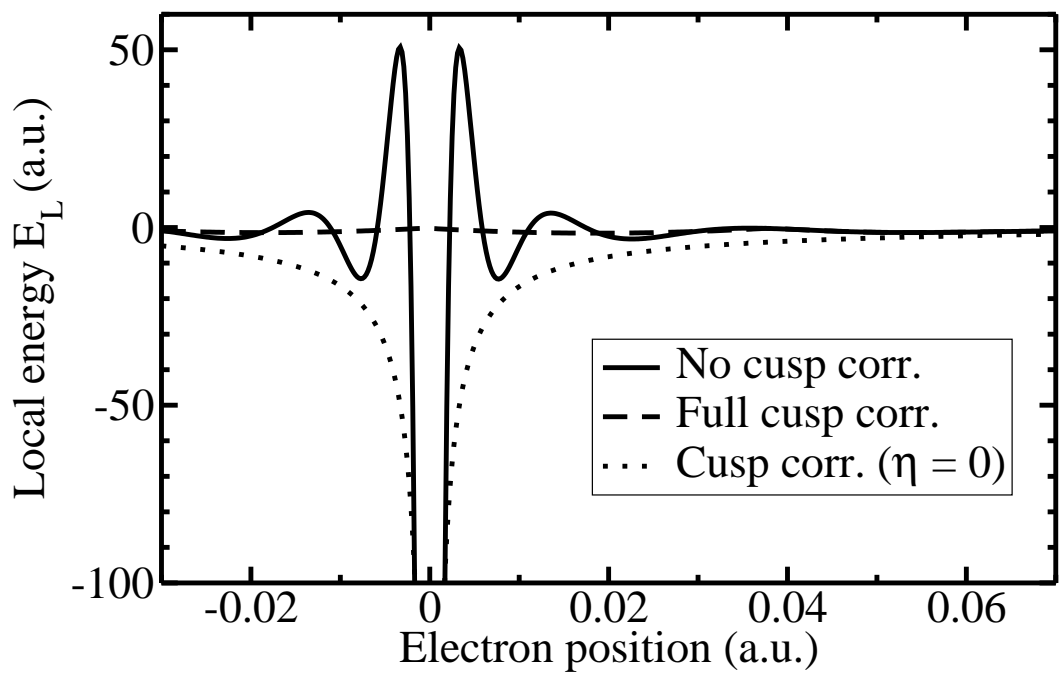}
\caption{The variation of the local energy, $E_L$, as an electron is
moved through one of the nuclei of a H$_2$ molecule of bond length
1.4\,a.u.  Slater-Jastrow wave functions are used, with orbitals which
are not cusp-corrected, orbitals which have the full cusp correction
imposed, and orbitals which have a partial cusp correction imposed for
which we set $\eta(0) = 0$.
\label{fig:EL_H2_bond1.4.eps}}
\end{center}
\end{figure*}

We have also tested our scheme for small molecules, in which the
contributions from the tails of the Gaussians centered on other sites
described by the $\eta$ term in Eq.~(\ref{eq:cusp_corrected}) are
significant. As a test case we studied the H$_2$ molecule, with a bond
length of $0.7395$\,\AA. We used an uncontracted Gaussian basis set
consisting of 11 $s$ functions and a single $p$ polarization function,
with all exponents optimized to minimize the energy. The final HF
energy obtained was $-1.128852$\,a.u. In
Fig.~\ref{fig:EL_H2_bond1.4.eps} we plot the local energy of the H$_2$
molecule calculated with the many-body Hamiltonian as one of the
electrons is moved through a nucleus.  Without the cusp correction the
local energy oscillates and diverges at the nucleus, but when the full
cusp correction is added the local energy is well behaved. To
understand the importance of including contributions from the tails of
Gaussians centered on other sites we have also plotted results with
$\eta(0)$ in Eq.~(\ref{eq:cusp_corrected}) artificially set to zero,
meaning that such contributions are not taken into account. (In fact,
$\eta(0) = 0.1879$ out of a total $\phi(0) = 0.9650$.)  It is apparent
from the figure that although the local energy does not oscillate, it
still diverges at the nucleus, demonstrating that one cannot satisfy
the cusp conditions exactly without taking into account basis function
contributions from other nuclei. This example demonstrates that our
cusp correction scheme completely removes the divergence in the local
energy when an electron moves through a nucleus, even in the
polyatomic case.

\subsection{Standard test set of small molecules}
\label{subsec:G2}

In order to demonstrate that our cusp-correction scheme gives a
general improvement across a wide range of chemical environments we
have performed VMC calculations with single-determinant wave functions
without a Jastrow factor (HFVMC calculations) for fifty-five molecules
taken from a standard benchmark set. Thirty-one of these molecules
were originally used to fit the semi-empirical G1 theory\cite{g1} with
a further twenty-four molecules containing elements from the second
row of the periodic table added to the set later\cite{g1a}.  We used
the standard molecular geometries specified for use with this set,
which were originally optimized at the MP2/6-31G($d$) level. We
emphasize that we used the automatic version of our algorithm with
$c_c=50$, and further optimization of the cusp correction radius was
not attempted.

In Table~\ref{table:g1_energies} we give results for the molecular HF
energies, $E_{\rm HF}$, calculated using the \textsc{crystal} code and
for the HFVMC energies, $E_{\rm HFVMC}$, (obtained with and without
cusp corrections) from the \textsc{casino} code. The HFVMC energies
were calculated from 250,000 samples in the electron configuration
space. The statistical error bars on the mean energies are given by
the number in brackets which represents the standard error in the last
digit. The table also shows the variance of the local energy,
$\sigma^2$, for each molecule, again with and without cusp
corrections. The error bars on the values of $\sigma^2$ obtained
without cusp corrections are very large, and the figure in brackets
represents the approximate standard error in the whole number.

\begin{figure*}
\begin{center}
\includegraphics[width=14cm,clip]{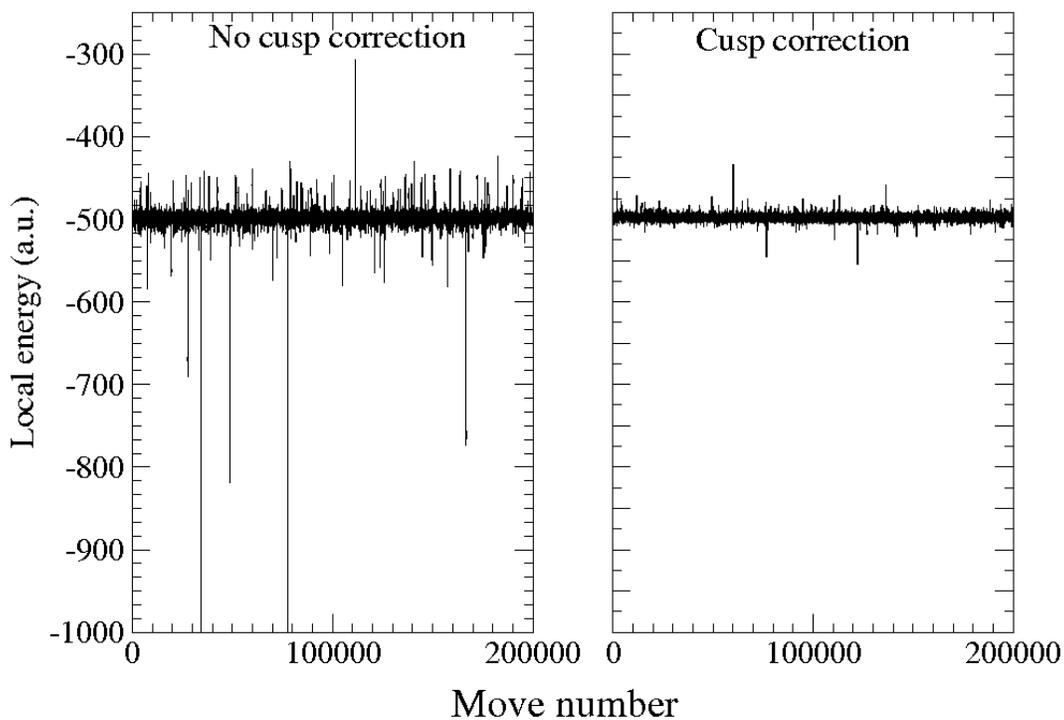}
\caption{Local energy after each VMC move for the CH$_3$Cl molecule
with and without cusp corrections.
\label{fig:vmcenergies.eps}}
\end{center}
\end{figure*}

It is clear that there is a general and very significant reduction in
the variance in the local energy for all fifty-five molecules on
introducing the cusp correction, with a consequent reduction in the
standard error in the mean energies by approximately an order of
magnitude. This is also apparent in Fig.~\ref{fig:vmcenergies.eps}
where the local energy is plotted as a function of move number for the
CH$_3$Cl molecule. In the absence of the cusp correction there are a
great many large spikes in the VMC energy resulting from the
divergences in the local energy near the nucleus. These are
significantly reduced when the orbitals are corrected.  We also found
that the correlation length for the energy was considerably reduced by
incorporating the cusp corrections.

In order to gauge the accuracy of all-electron quantum Monte Carlo and
of our cusp-correction scheme, we have performed benchmark
calculations of the DMC energies of this set of molecules and their
constituent atoms, and hence the molecular atomization energies and
their mean absolute deviation from experiment. Details of these
calculations will appear in a separate publication\cite{me}, together
with a comparison with the results of Grossman\cite{grossman_2002} who
performed similar DMC calculations using pseudopotentials.

\section{Conclusions}

We have described and tested a simple, automatic, numerically-stable
scheme for introducing the correct cusp at the nucleus into orbitals
obtained from calculations using Gaussian basis sets.  This ensures
that the local energy is finite when an electron and nucleus are
coincident.  Our scheme may readily be adapted for use with other
atom-centered basis sets.

The scheme has been devised for use within all-electron VMC and DMC
calculations.  We have performed extensive tests for the Ne atom, the
H$_2$ molecule and a fifty-five molecule benchmark set. In all cases
it greatly reduces the variance of the energy and also slightly
reduces the variational energy. This technical development should lead
to improved results from all-electron VMC and DMC calculations.

\begin{table*}
\begin{center}
\begin{tabular}{lccccc}
\hline \hline

Molecule & $E_{\rm HF}$ (a.u.) & $E_{\rm HFVMC}$ (a.u.) & $\sigma^2$ (a.u.) & $E_{\rm HFVMC}$ (a.u.) & $\sigma^2$ (a.u.) \\
& & (no cusp correction)& & (cusp correction)&\\

\hline

BeH             & -15.1519  & -15.153(5) & 19(4)        & -15.1520(7) &   3.04(4) \\    
C$_2$H$_2$      & -76.8435  & -76.86(2)  & 236(64)      & -76.850(2)  &  17.2(2)  \\ 
C$_2$H$_4$      & -78.0602  & -78.13(3)  & 299(67)      & -78.065(2)  &  17.6(2)  \\ 
C$_2$H$_6$      & -79.2567  & -79.24(2)  & 156(31)      & -79.259(2)  &  18.0(2)  \\ 
CH              & -38.2809  & -38.31(1)  & 160(48)      & -38.284(1)  &   8.2(1)  \\ 
CH$_2$ singlet  & -38.8914  & -38.90(2)  & 252(139)     & -38.891(2)  &   8.5(1)  \\
CH$_2$ triplet  & -38.9184  & -38.91(1)  & 87(31)       & -38.920(2)  &   8.2(1)  \\
CH$_3$          & -39.5761  & -39.559(8) & 46(9)        & -39.581(2)  &   8.3(1)  \\ 
CH$_3$Cl        & -499.1365 & -499.3(2)  & 13770(12046) & -499.121(9) & 148(4)    \\ 
CH$_4$          & -40.2120  & -40.22(2)  & 178(100)     & -40.215(2)  &   8.9(1)  \\ 
Cl$_2$          & -918.9768 & -919.0(1)  & 3850(1297)   & -918.97(1)  & 270(2)    \\ 
ClF             & -558.8792 & -558.72(6) & 1472(218)    & -558.867(9) & 164(3)    \\ 
ClO             & -534.2931 & -534.3(1)  & 2768(789)    & -534.311(7) & 150(1)    \\ 
CN              & -92.2325  & -92.21(2)  & 229(45)      & -92.235(2)  &  20.4(2)  \\ 
CO              & -112.7699 & -112.75(2) & 315(62)      & -112.773(3) &  27.0(4)  \\ 
CO$_2$          & -187.6880 & -187.62(2) & 384(39)      & -187.691(3) &  44.9(5)  \\ 
CS              & -435.3424 & -435.17(5) & 868(120)     & -435.352(7) & 121(2)    \\ 
F$_2$           & -198.7482 & -198.73(3) & 736(174)     & -198.749(4) &  51.3(8)  \\ 
H$_2$CO         & -113.9042 & -114.04(8) & 1541(922)    & -113.909(3) &  26.8(3)  \\ 
H$_2$O          & -76.0551  & -76.04(2)  & 188(72)      & -76.054(2)  &  18.4(3)  \\ 
H$_2$O$_2$      & -150.8311 & -150.85(4) & 800(267)     & -150.835(3) &  35.9(4)  \\ 
H$_2$S          & -398.7057 & -398.68(6) & 2355(1534)   & -398.703(7) & 113(1)    \\ 
H$_3$COH        & -115.0837 & -115.14(4) & 1113(672)    & -115.088(3) &  27.2(3)  \\ 
H$_3$CSH        & -437.7466 & -437.81(9) & 1796(678)    & -437.766(8) & 121(1)    \\ 
HCl             & -460.0976 & -459.87(6) & 1437(427)    & -460.109(7) & 135(2)    \\ 
HCN             & -92.9004  & -92.92(2)  & 195(22)      & -92.903(2)  &  21.0(3)  \\ 
HCO             & -113.2845 & -113.28(2) & 365(105)     & -113.289(3) &  26.7(5)  \\ 
HF              & -100.0541 & -100.03(3) & 477(200)     & -100.050(3) &  26.3(5)  \\ 
HOCl            & -534.9018 & -534.86(6) & 1301(141)    & -534.901(8) & 152(2)    \\ 
Li$_2$          & -14.8703  & -14.874(3) & 23(5)        & -14.8711(7) &   3.06(3) \\ 
LiF             & -106.9778 & -106.95(2) & 470(199)     & -106.982(3) &  26.7(3)  \\ 
LiH             & -7.9859   & -7.984(2)  & 6.7(9)       & -7.9864(5)  &   1.64(3) \\    
N$_2$           & -108.9710 & -108.96(2) & 308(76)      & -108.973(3) &  25.1(3)  \\ 
N$_2$H$_4$      & -111.2203 & -111.17(2) & 180(22)      & -111.224(3) &  26.3(6)  \\ 
Na$_2$          & -323.6914 & -323.67(4) & 868(165)     & -323.697(5) &  87(2)    \\ 
NaCl            & -621.4350 & -621.4(1)  & 4278(1945)   & -621.425(8) & 187(6)    \\ 
NH              & -54.9798  & -54.99(3)  & 343(218)     & -54.982(2)  &  12.5(5)  \\ 
NH$_2$          & -55.5849  & -55.54(1)  & 82(21)       & -55.588(2)  &  12.4(2)  \\ 
NH$_3$          & -56.2173  & -56.17(1)  & 75(11)       & -56.222(2)  &  12.7(2)  \\ 
NO              & -129.2943 & -129.30(3) & 442(91)      & -129.297(3) &  30.8(5)  \\ 
O$_2$           & -149.6574 & -149.64(3) & 520(123)     & -149.663(3) &  36(1)    \\ 
OH              & -75.4188  & -75.45(2)  & 319(102)     & -75.421(2)  &  18.0(3)  \\ 
P$_2$           & -681.4717 & -681.42(7) & 3204(1420)   & -681.482(9) & 195(4)    \\ 
PH$_2$          & -341.8802 & -341.88(6) & 1099(408)    & -341.885(6) &  99(3)    \\
PH$_3$          & -342.4814 & -342.41(8) & 1486(855)    & -342.487(7) &  97(1)    \\ 
S$_2$           & -795.0756 & -795.1(1)  & 8525(6227)   & -795.081(9) & 228(3)    \\ 
Si$_2$          & -577.5901 & -577.54(8) & 2604(889)    & -577.604(7) & 157(1)    \\ 
Si$_2$H$_6$     & -581.3623 & -581.28(5) & 1135(142)    & -581.377(8) & 159(2)    \\ 
SiH$_2$ singlet & -290.0261 & -289.93(4) & 781(191)     & -290.031(6) &  82(2)    \\
SiH$_2$ triplet & -290.0047 & -289.95(5) & 1348(571)    & -290.001(6) &  79(1)    \\
SiH$_3$         & -290.6362 & -290.66(6) & 1352(506)    & -290.629(6) &  80(1)    \\
SiH$_4$         & -291.2569 & -291.27(7) & 1046(332)    & -291.251(6) &  82(1)    \\ 
SiO             & -363.8279 & -363.87(8) & 3688(2648)   & -363.835(6) &  97(1)    \\ 
SO              & -472.3826 & -472.26(5) & 1107(103)    & -472.392(7) & 129(1)    \\ 
SO$_2$          & -547.2624 & -547.3(1)  & 5672(3324)   & -547.286(7) & 148(2)    \\
\hline \hline
\end{tabular}
\caption{HF energies and, for the VMC calculations, the variance
$\sigma^2$, for the test set of fifty-five molecules.
\label{table:g1_energies}}
\end{center}
\end{table*}

\section{Acknowledgments}

We acknowledge financial support from the Engineering and Physical
Sciences Research Council (EPSRC) of the United Kingdom.


\begin{thebibliography}{}
\bibitem{footnote_units} We use Hartree atomic units,
$\hbar=|e|=m_e=4\pi \epsilon_0=1$, throughout this article.
\end{thebibliography}
\end{document}